\documentstyle[12pt,epsf]{article}
\begin{document}

\title{On controlling simple dynamics by a disagreement function.}
\author{Katarzyna Sznajd-Weron\\
Institute of Theoretical Physics, University of Wroc{\l}aw}
\date{\today}
\maketitle

We introduce a formula for the disagreement function which is used to control a recently 
proposed dynamics of the Ising spin system. This leads to four different phases
of the Ising spin chain in a zero temperature. One of these phases is doubly degenerated
(anti- and ferromagnetic states are equally probable).
On the borders between the phases two types of transitions are observed: 
infinite degeneration and instability lines.
The relaxation of the system depends strongly on the phase. 

\vspace{1cm}

\section{Introduction}
The Ising spin system is one of the most frequently used models of statistical mechanics.
Its simplicity (binary variables) makes it appealing to researchers from other branches
of science including biology \cite{DH91}, sociology \cite{SH00} and economy 
\cite{CS99,SMR00}. 
In sociophysics models of opinion formation based on the social impact theory (reviewed
in \cite{HKS01}), the individual opinion is decribed by the Ising spin. This corresponds not only 
to typical "yes"-"no" questions, but also to important issues where the distribution of 
opinion seems to be bimodal, peaked on extreme values. In general, in these models the influence 
flows inward from the border to the center, like in the majority rules, where the site in the
middle takes the state of the majotiry of neighbouring sites. In contrast,
in our (USDF) model \cite{SWS00} (reviewed in \cite{S02}) an outward flow of influence is imposed. 
In the USDF model, an isolated person does not convince others; however, a group of people 
sharing the same opinion influences their neighbors. In spite of simple rules 
the model exhibited complicated dynamics in one \cite{SWS00} and more (references in
\cite{S02}) dimensions. In less than a year, this model 
has found several applications: e.g. it was used to explain the distribution of votes among 
candidates in Brazilian local election \cite{BSK02} and to model the price dynamics
\cite{SWW02}. 

In this paper we introduce the 
{\it "disagreement function"} \cite{CS99} which is used control the dynamics of the model.
We show that for a one dimensional Ising spin chain in zero temperature this leads to 
four different phases: ferromagnetic, antiferromagnetic, (2,2) antiphase and a doubly degenerated
phase in which both ferromagnet and antiferromagnet are equally probable stable steady states 
of the system.   
Apart of structural differences between phases the difference in relaxation will be shown. 
The system in general will relax in two different ways depending on the phase.
Moreover, a sharp change of the relaxation time on borders of the phases will be observed.

\section{The model}
Recently a simple model for opinion evolution in closed community was proposed \cite{SWS00}.
In this model the community is represented by a horizontal chain of Ising spins, which
are either up or down. A pair of parallel neighbors forces its two neighbors to have the same
orientation (in random sequential updating), while for an antiparallel pair, the left neighbor 
takes the orientation ot the right part of the pair, and left neighbor follows the right part
of the pair. 
Thus the model can be described by two simple dynamic rules:
\begin{itemize}
\item
$D_1$: $S_{i-1}(t+1)=S_{i}(t)$ and $S_{i+2}(t+1)=S_{i}(t)$ if $S_{i}(t)*S_{i+1}=1$
\item
$D_2$: $S_{i-1}(t+1)=S_{i+1}(t)$ and $S_{i+2}(t+1)=S_{i}(t)$ if $S_{i}(t)*S_{i+1}=-1$.
\end{itemize}
In contrast to usual majority rules \cite{A91}, in this model the influence does
not flow inward from the surrounding neighbors to the center site, but spreads outward from
the center to the neighbors. The model thus describes the spread of opinions. The dynamic rules
leads to two different stable steady states (ferromagnetic and antiferromagnetic) with equal
probability. The second dynamic rule ($D_2$) of the model has been already changed 
in two different ways. In case of antiparallel spins the neighboring spins can either flip 
with probability 1/2 \cite{SWW02} ($D_{2A}$) or remain unchanged \cite{S02}
($D_{2B}$). In both cases ($D_{2A}$ and $D_{2B}$) the only final state is ferromagnet. 
It is worth to mention that the ferromagnetic state for both rules, $D_{2A}$ and $D_{2B}$, 
is always reached (even in two dimensions) in contrast to the Ising spin system under 
Glauber dynamics \cite{MAS01,SKR01}. In the case of $D_{2B}$ besides of ferromagnetic stable
steady states, the antiferromagnetic unstable steady state exsits. 

Since, we have up till now three different rules for the case of antiparallel spins, we 
propose a generalization of the previous models. 
The generalized model consits of two components, hence the name TC model:
\begin{itemize}
\item
The dynamics: choose a pair of spins $S_{i+1}$ and $S_{i+2}$ and change its next nearest
neighbors $S_{i}$ and $S_{i+3}$
\item
The rules: control the dynamics of the $i$-th and $(i+3)$-th spins by the disagreement function.
\end{itemize} 

In the next sections we introduce the disagreement function and show that TC model
includes as a special cases all earlier proposed models \cite{SWS00,SWW02,S02}.
Moreover, TC model consits of more then those three subcases which we present on
its phase diagram. Using Monte Carlo simulations we show how the system described
by TC model relax.

\section{How to control dynamics?}
Let us assume for a while that we have the formula for a function that can control 
TC dynamics and denote it by $E$. We choose at random a pair of spins $S_{i+1}$ and $S_{i+2}$ 
and we calculate $E^+=E(S_{i},S_{i+1},S_{i+2})$. Next we calculate $E^-=E(-S_{i},S_{i+1},S_{i+2})$ 
in the case of flipped $i$-th spin. 
If $E^-<E^+$ then we will flip the $i$-th spin, if not the spin will remain unchanged. 
We do the same for the second neighbor of the choosen pair i.e. for the spin $S_{i+3}$. 

Our dynamics looks now similar to the Glauber dynamics in zero temperature, 
where $E$ plays the role of energy. However, there are three main differences between 
these two dynamics:
\begin{itemize}
\item
In Glauber dynamics we flip the $i$-th spin according to the interactions with $(i-1)$-th 
and $(i+1)$-th spins, here we look at $(i+1)$-th and $(i+2)$-th spins. 
\item
In Glauber dynamics the flip is done even if the old energy is equal to the new one. It doesn't
seem natural in zero temperature but it is needed to get the ground state (in two
dimensions even this is not enough \cite{SKR01}).
\item
In our case $E$ is called the disagreement function, since it is not the energy.
\end{itemize}
Now we will look for the formula for $E$.
We shall deal with the lattice model where each lattice site $i$ is occupied by an ising spin 
$S_i= \pm 1$. Usually, the spins are assumed to interact through pairwise coupling of the form
$-J_{ij}S_iS_j$, where $J_{ij}$ are exchange integrals. Of course, the ordering of the spins
is determined by the interactions. One of the best studied examples is the nearest neighbour 
(nn) Ising model with ferromagnetic coupling, i.e. $J_{ij}=J>0$ for neighbour spins $S_i$
and $S_j$, while $J_{ij}=0$ for more distant spins. Certainly, in a such model, the spins form 
the ferromagnetic state (all spins up or all spins down) in low temperature. 
For $J<0$ the antyferromagnetic state is formed in low temperture.

In TC model the $i$-th spin interacts with its two neighbors, and the
1D hamiltonian can be written in the following form:
\begin{equation}
H=-J_1 \sum_i S_i S_{i+1} - J_2 \sum_i S_i S_{i+2}.
\label{annni}
\end{equation}
For $J_1>0$ and $J_2<0$ this is the well known ANNNI (axial next-nearest neighbour Ising) model 
introduced in \cite{FS80} and reviewed in \cite{S88}.
It describes the Ising spin chain with ferromagnetic interaction $J_1>0$ between
nearest neighbours (nn) and antiferromagnetic interactions between next nearest neighbours (nnn).
Of course, in the one-dimensional case truly ordered states are stable only in zero temperature
$T=0$. If we introduce the competition ratio $r=-J_2/J_1$ we get in $T=0$ ferromagnetic state
for $r<1/2$ and (2,2) structure for $r>1/2$. 

Now, we will use the nnn Ising hamiltonian [\ref{annni}] to construct the dissagrement function $E$. Chowdbury and Stauffer introduced similary a dissagrement function based on the simple nn Ising hamiltonian to the model of finantial market \cite{CS99}. 
We write $E$ in the following form:
\begin{equation}
E=-J_1S_i S_{i+1} - J_2S_i S_{i+2}.
\end{equation}
Each individual would like to minimize the corresponding disagreement function.
In the TC dynamics we choose a pair $S_{i+1}$ and $S_{i+2}$ and we change its 
neighbour $S_{i}$ (we also change $S_{i+3}$ spin calculating 
$E=-J_1S_{i+3} S_{i+2} - J_2S_{i+3} S_{i+1}$, but for simplicity we further write only about
$i$-th spin). 
For these three spins ($S_i, S_{i+1}, S_{i+2}$) we have four values of $E$:
\begin{enumerate}
\item
$+ + +$ or $- - -$ gives $E_1=-(J_1+J_2)$ 
\item
$- + +$ or $+ - -$ gives $E_2=J_1+J_2$ 
\item
$+ - +$ or $- + -$ gives $E_3=J_1-J_2$ 
\item
$- - +$ or $+ + -$ gives $E_4=J_2-J_1$ 
\end{enumerate}

It is worth to notice that the possible transitions are only between 
states 1 and 2 or between 3 and 4. Now we can derive from the TC model all previous models:
\begin{itemize}
\item
USDF model \cite{SWS00}:\\
if $S_{i+1}(t)*S_{i+2}(t)=1$ then $S_{i}(t+1)=S_{i+1}(t)$
i.e. $E_1<E_2$ \\
if $S_{i+1}(t)*S_{i+2}(t)=-1$ then $S_{i}(t+1)=S_{i+2}(t)$
i.e. $E_3<E_4$. \\ 
Thus USDF model correspond to the TC model with  $-J_2<J_1<J_2$.
\item
The model of financial market \cite{SWW02}:\\
if $S_{i+1}(t)*S_{i+2}(t)=1$ then $S_{i}(t+1)=S_{i+1}(t)$\\
if $S_{i+1}(t)*S_{i+2}(t)=-1$ then $S_{i}(t+1)=-S_{i}(t)$ with probability $1/2$.\\
This corresponds to the TC model with $E_1<E_2$ and $E_4<E_3$ 
$\Rightarrow -J_2<J_1$ and $J_1>J_2$.
\item
Other models reviewed in \cite{S02}:\\ 
if $S_{i+1}(t)*S_{i+2}(t)=1$ then $S_{i}(t+1)=S_{i+1}(t)$
i.e. $E_1<E_2$ \\
if $S_{i+1}(t)*S_{i+2}(t)=-1$ then $S_{i}(t+1)=S_{i}(t)$
i.e. $E_3=E_4$ \\
These models correspond to the TC model with $J_1=J_2$.
\end{itemize} 
There are of course more subcases of the TC model depending on interaction coefficients
$J_1$ and $J_2$. On the Figure 1 all possible phases, depending on interaction 
coefficients, are presented. The North (doubly degenerated) phase corresponds to the original
rule $D_2$. The East (ferromagnetic) phase corresponds to rule $D_{2A}$ 
(the flip in case of antiparallel spins is made at random). 
The line between these two phases corresponds to rule $D_{2B}$ (the flip is possible only 
in case of parallel spins).
On this line the antiferromagnetic steady state still exists but it becomes unstable and we
never reach it outside of this state.
It is also interesting to see what happens on other border lines. 
The border between the ferromagnetic state and the (2,2) antiphase (see Fig.2) 
is infinitely degenerated. Let us define (after \cite{S88}) 
a $k$-band formed by $k$ adjacent, identically oriented spins, terminated at the both ends by
opposite oriented spins. With such a definition, the ferromagnetic structure is zero-band,
antiferromagnetic is one-band and (2,2) antiphase is a two-band structure. On the line between
ferromagnet and (2,2) antiphase any sequence of $k$-band ($k \ge 2$) is equally probable
(see Fig.2). The line between (2,2) antiphase and antiferromagnet is also degenerated, 
and  any sequence of k-band (with k=1,2) is the steady state (see Fig.2). 

There is also another interesting feature which differs phases from each other - 
the time and the style in which the system relax. We will describe it in the next section.

\section{How does the system relax?}
What happens when we suddenly cool our system from a high temperature to zero temperature?
As we mentioned previously the system will relax to one of the possible final states described
by the phase diagram (Fig.1). But how does it relax? We studied this using Monte Carlo
simulations. We found out that the relaxation process
strongly depends on phase. The system can reach antiferromagnetic state in the West 
(antiferromagnetic) phase as well as in the North (degenerated)  phase.
However, it will relax to this state differently in each case. 
In the antiferromagnetic phase the system will be almost totally ordered after several Monte Carlo
Steps (MCS). Then the system will oscillate oround the final state. These oscillations will
decrease in time and finally the system will reach the steady state. In the degenerated phase
the system will order very slowly.

In Figure 3 the examples of relaxations in all four phases are presented. 
To show this relaxation we choose the opinion changes,
since the model was proposed to investigate the opinion dynamics.
We defined the opinion \cite{SWS00} as a magnetization of the system:
\begin{equation} 
m=\sum_{i=1}^N S_i,
\end{equation} 
For such a choice the system will relax to $|m|=1$ (ferromagnet) or $|m|=0$ (antiferromagnet
or (2,2) antiphase). Of course, one could also choose the two point correlation function
$g=<S_iS_{i+1}>$ to see how the system relaxes. We have done it to recognize the
final state ($g=1,-1$ or $0$ for ferromagnet, antiferromagnet and (2,2) antiphase 
respectively).
For $J_1>-J_2$ (North and East part of the diagram in Fig.1) the ordering of the system 
is very slow. Sometimes the opinion can change dramaticaly in a short time (see Fig.3). 
The long time trends are observed, which reminds very much of the real sociological procesess
\cite{SWS00}. For $J_1<-J_2$
the system is almost ordered after several Monte Carlo steps, however, then it takes a
long time to reach the real final steady state. The opinion is fluctuating around zero and
these fluctuations are decreasing in time (see Fig.3). 
Although the way in which the system relaxes in the North and East phases is the same,
the relaxation time in each of these phases is different. About a two times shorter (in average) 
time is needed to reach the final state in the degenerated phase. 
The relaxation time changes very sharply
on the border between these two phases (Fig.4). 
A similar effect is observed also on the border between the antiferromagnetic and degenerated 
phases.

\section{Summary}
We proposed the new generalized model of opinion formation.
The disagrement function was introduced to control the simple dynamics of an Ising spin chain
in zero temperature. This allowed to generalize the previous model of opinion dynamics.
It was shown that the phase diagram for that system described by such a model 
consists of four different
phases. The most interesting is the existence of the doubly degenerated phase in which the
system can reach the antiferromagnetic steady state or the ferromagnetic steady state with the same 
probability. Moreover, it was shown that the system can relax in two different ways depending
on the interaction coefficients. Surprisingly the system can reach the antiferromagnetic state in two 
different ways. In the antiferromagnetic phase the system will be almost ordered after several
Monte Carlo steps and then decreasing oscillations around the final state will lead the 
system into this state. In the degenerated phase, the system will behave 
"blindly" making a long "random" walk to the final state. 
It would be probably worth to look at the system described by such a model in
higher dimensions and higher temperature.
We also hope that the generalized TC model will find so many applications 
as its older brothers \cite{SWS00}. 

\vspace{1cm}
I would like to thank the Fundation for Polish Science (FNP) for the finantial support.

\begin{figure}[p]
\centerline{\epsfxsize=10cm \epsfbox{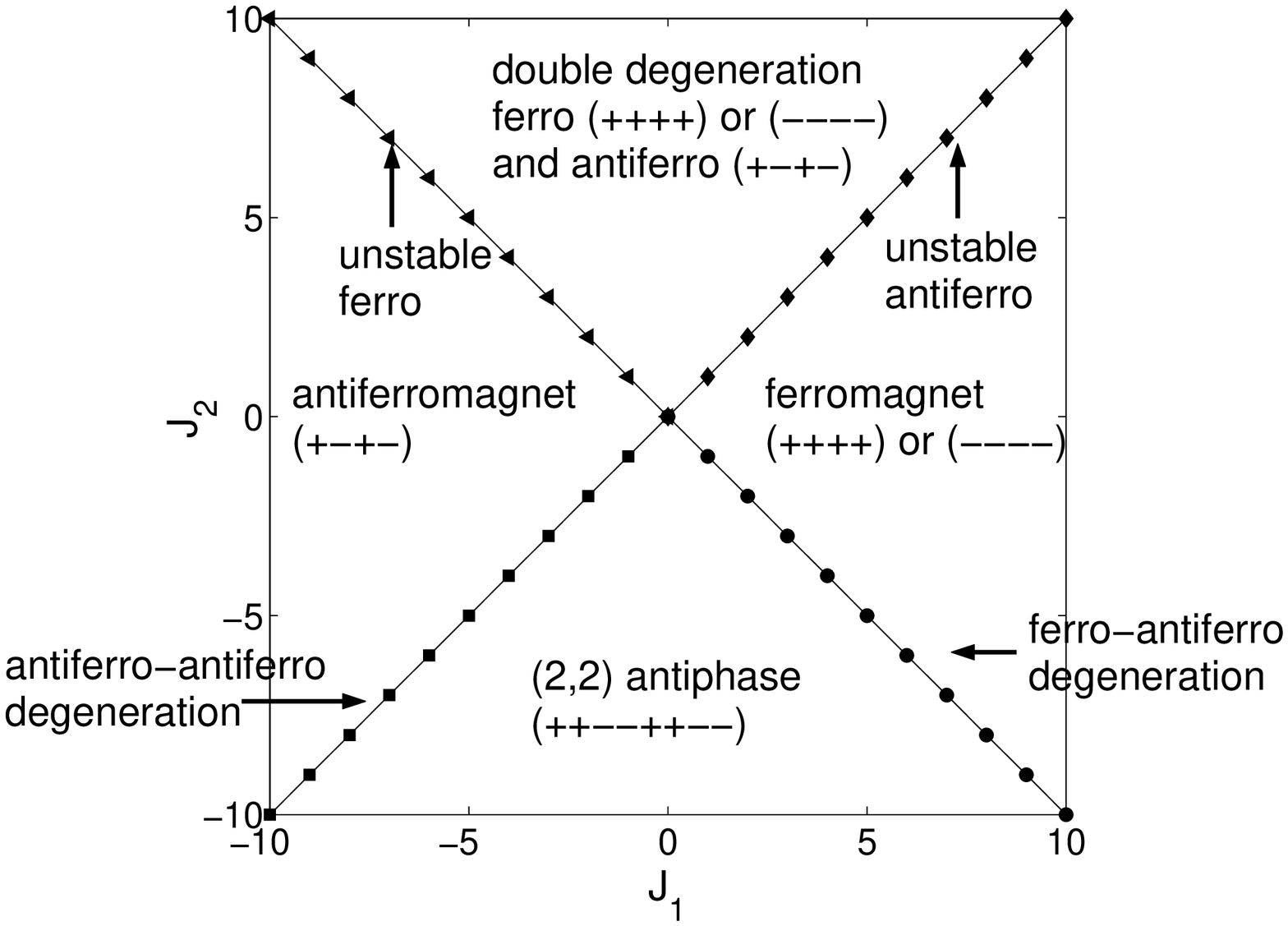}}
\caption{The phase diagram of the TC model}. 
\end{figure}

\begin{figure}[p]
\centerline{\epsfxsize=10cm \epsfbox{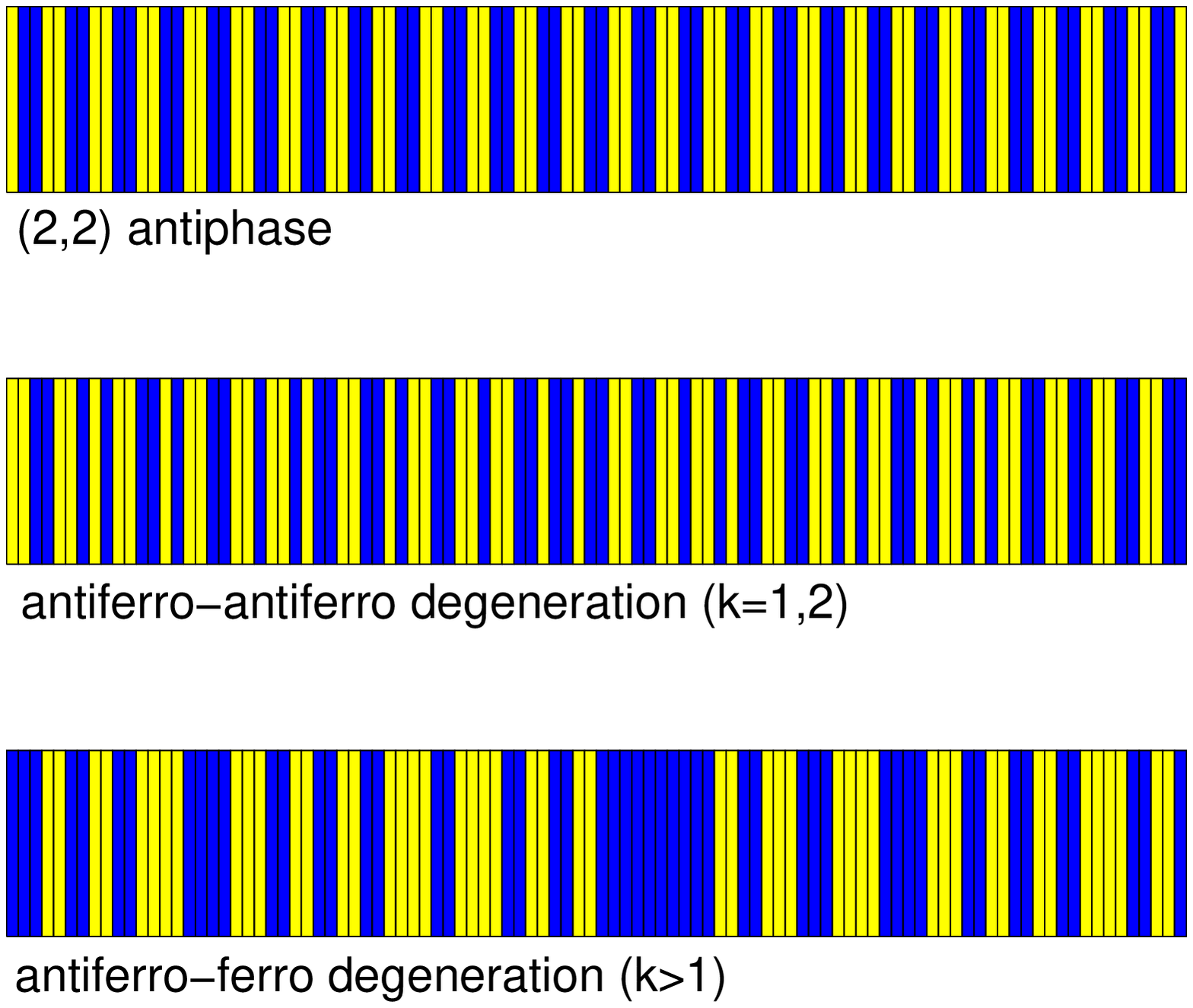}}
\caption{Examples of 3different steady states of the TC model are presented. 
Bright lines denote spins up and dark lines denote spins down.}
\end{figure}

\begin{figure}[p]
\centerline{\epsfxsize=10cm \epsfbox{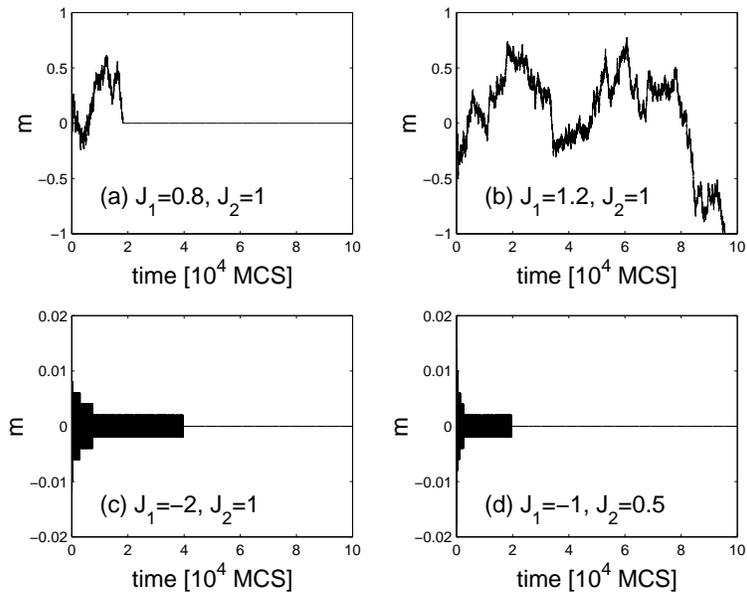}}
\caption{Examples of the relaxation for 1000 spins system are presented.
Two kinds of relaxations where observed depending on interaction coefficients.
For $J_1>-J_2$ the system makes long "random" walk to the final state, while
for $J_1<-J_2$ the system makes decreasing oscillations around the final state.}
\end{figure}

\begin{figure}[p]
\centerline{\epsfxsize=10cm \epsfbox{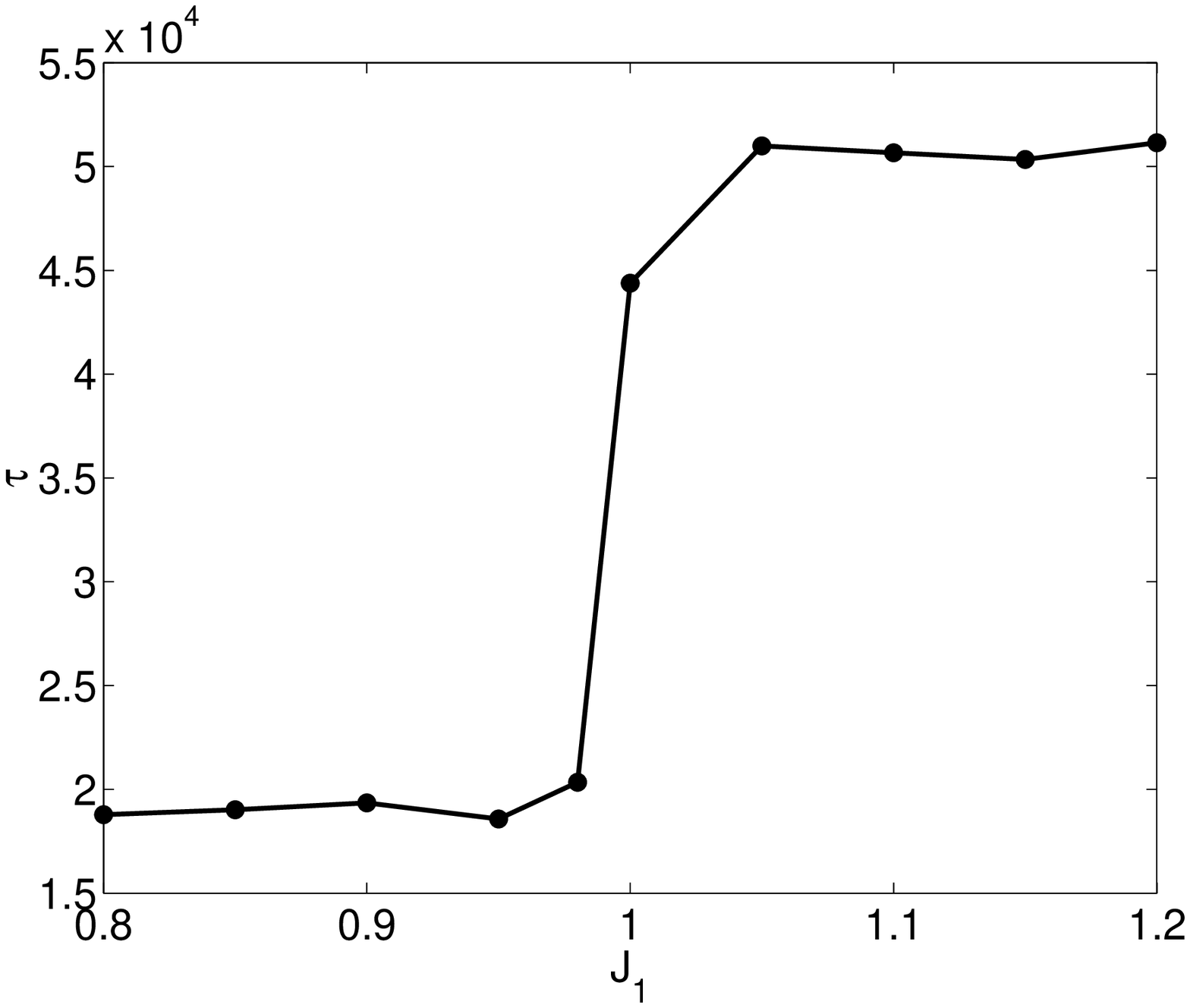}}
\caption{Relaxation time for $J_2=1$. On this figure we present results for the system
of 1000 spins averaged over 10000 samples.}
\end{figure}


\begin{thebibliography}{33}
\bibitem{DH91}
D. Derrida and P. G. Higgs, J. Phys. A {\bf 24}, L985 (1991)
\bibitem{SH00}
F. Schweitzer and J. A. Ho{\l}yst, Eur. Phys. J. B {\bf 15}, 723 (2000)
\bibitem{CS99}
D. Chowdbury and D. Stauffer, Eur. Phys. J. B {\bf 8}, 477 (1999)\bibitem{SMR00}
R. Savit, R. Manuca, and R. Riolo, Phys. rev. Lett. {\bf 82}, 2203 (1999)
\bibitem{HKS01}
J. A. Ho{\l}yst, K. Kacperski and F. Schweitzer, Annual Review of Computational Physics 
IX, p.275, World Scientific, Singapore 2001 
\bibitem{SWS00}
K. Sznajd-Weron and J. Sznajd, Int. J. Mod. Phys. C {\bf 11}, 1157 (2000)
\bibitem{S02}
D. Stauffer, Journal of Artificial Societies and Social Simulation vol. 5, no.1
(2002)
\bibitem{BSK02}
A. T. Bernardes, D. Stauffer and K. Kert{\'e}sz, Eur. Phys. J. B {\bf 25} 123 (2002)
\bibitem{SWW02}
K. Sznajd-Weron and R. Weron, Int. J. Mod. Phys. C {\bf 13}, 115 (2002)
\bibitem{A91}
J. Adler, Physica A {\bf 171}, 453 (1991)
\bibitem{MAS01}
A. A. Moreira, J. S. Andrade Jr. and D. Stauffer, Int. J. Mod. Phys. C {\bf 12}, 39 (2001)
\bibitem{SKR01}
V. Spirin, P. L. Krapivsky and S. Redner, Phys. Rev. E {\bf 63}, 036118 (2001)
\bibitem{FS80}
M. E. Fisher and W. Selke, Phys. Rev. Lett. $\bf 44$ 1502 (1980)
\bibitem{S88}
W. Selke, Phys. Rep. 170, 213 (1988) 
\end{thebibliography}
\end{document}